\documentstyle[12pt,fullpage,epsfig]{article}

\newcommand{\fig}[3] 
{%
 \begin{figure}[t]
 \begin{center}
 \input{#1}
 \end{center}
 \caption{#3}
 \label{#2}
 \end{figure}
}

\newtheorem{theorem}{Theorem}[section]
\newtheorem{lemma}[theorem]{Lemma}
\newtheorem{corollary}[theorem]{Corollary}
\newtheorem{proposition}[theorem]{Proposition}

\newtheorem{definitions}{Definitions}

\newcommand{\Opt}[1]{\mbox{${\cal OPT}(#1)$}}
\newcommand{\MaxCycle}[1]{{\cal C}}

\newcounter{algorithmLine}

\newenvironment{proof}{\par\noindent {\em Proof. }}{\myendproof}
\def\myendproof{\hfill{\vbox{\hrule\hbox{%
   \vrule height1.3ex\hskip0.8ex\vrule}\hrule}}}

\title{\large\bf On Strongly Connected Digraphs with Bounded Cycle Length}

\author{Samir Khuller
  \thanks{Computer Science Department and Institute for Advanced Computer
    Studies, University of Maryland, College Park, MD~20742.
    Research supported by NSF Research Initiation Award CCR-9307462.
    E-mail~: {\tt samir@cs.umd.edu}.}
  \and Balaji Raghavachari
  \thanks{Department of Computer Science, The University of Texas at Dallas,
    Box 830688, Richardson, TX 75083.
    Research supported by NSF grant CCR-9409625.
    E-mail : {\tt rbk@utdallas.edu}.}
  \and Neal Young
  \thanks{Corresponding author.
    Department of Computer Science, Dartmouth College, Hanover, NH 03755,
    USA. E-mail: neal.young@dartmouth.edu.
    Part of this research was done while at
    School of ORIE, Cornell University, Ithaca NY 14853
    and supported by \'Eva Tardos' NSF PYI grant DDM-9157199.
  }
}

\date{ }

\begin{document}

\maketitle

\begin{abstract}
  Given a directed graph $G=(V,E)$, a natural problem
  is to choose a minimum number of the edges in $E$
  such that, for any two vertices $u$ and $v$,
  if there is a path from $u$ to $v$ in $E$,
  then there is a path from $u$ to $v$ among the chosen edges.
  We show that in graphs having no directed cycle with more than three edges,
  this problem is equivalent to Maximum Bipartite Matching.
  This leads to a small improvement in the performance guarantee
  of the  previous best approximation algorithm for the general problem.
\end{abstract}
\section{Introduction}
Let $G=(V,E)$ be a directed graph. 
The minimum equivalent graph (MEG) problem is the following:
find a smallest subset $S\subseteq E$ of the edges
such that, for any two vertices $u$ and $v$,
if there is a path from $u$ to $v$ in $E$
then there is a path from $u$ to $v$ using only edges in $S$.
The problem is NP-hard~\cite{GJ}.
A $c$-approximate solution is a subset of edges
providing the necessary paths
of size at most $c$ times the minimum.
A $c$-approximation algorithm is a polynomial-time algorithm
guaranteeing a $c$-approximate solution.

Moyles and Thompson \cite{MT} observed that
any solution to the MEG problem 
decomposes into solutions for each strongly connected component
and a solution for the component graph
(the graph obtained by contracting each strongly connected component).
Thus, the problem reduces in linear time to two cases:
the graph is either acyclic or strongly connected.
If the graph is acyclic, the MEG problem
is equivalent to the {\em transitive reduction}\/ problem,
which was shown by Aho, Garey and Ullman
to be equivalent to transitive closure~\cite{AGU}.
Thus, we assume the graph is strongly connected,
so that the problem is to find a small subset of the edges
preserving the strong connectivity.
We refer to this problem as
{\em the strongly connected spanning subgraph (SCSS)}\/ problem.

The only known $c$-approximation algorithm for any $c<2$
works by repeatedly contracting cycles \cite{KRY-JOURNAL}.
Each cycle contracted is either a longest cycle in the current graph,
or has length at least some constant $k$.
The set of contracted edges yields the set $S$.
As $k$ grows, the performance guarantee of this algorithm
rapidly tends to $\pi^2/6\approx 1.64$.

A natural modification is to solve the problem optimally
as soon as the maximum cycle length in the current graph
drops below some threshold.
The problem remains NP-hard even when the maximum cycle length is five,
but we conjectured in \cite{KRY-JOURNAL}
that it was solvable in polynomial time
if the maximum cycle length is three.
We use SCSS$_3$ to denote the SCSS problem with this restriction.
In this paper we confirm the conjecture:
\begin{theorem} \label{main-result}
  The SCCS$_3$ problem in $n$-vertex digraphs
  reduces in $O(n^2)$ time to Minimum Bipartite Edge Cover.
\end{theorem}
This gives an $O(n^2+m\sqrt{n})$-time algorithm for the SCCS$_3$ problem,
since Minimum Bipartite Edge Cover is trivially equivalent
to Maximum Bipartite Matching~\cite{NR},
which can be solved in $O(m\sqrt{n})$ time \cite{HK}.
Modifying the cycle-contraction algorithm correspondingly
reduces its performance guarantee by $1/36$:
\begin{corollary} \label{corollary}
  For any $c > \pi^2/6 - 1/36 \approx 1.61$,
  there exists a $c$-approximation algorithm for the MEG problem.
\end{corollary}
(For graphs with bounded cycle size,
a slightly stronger performance guarantee can be shown
as described at the end of Section~\ref{appl_sec}.)
This corollary follows from a straightforward modification
to the analysis in \cite{KRY-JOURNAL} of the algorithms described above.

Here is an overview of the reduction of SCSS$_3$ to Edge Cover.
We classify each edge as either {\em necessary}\/
(removal of the edge leaves the graph not strongly connected)
or {\em redundant}\/ (otherwise).
It turns out that any SCSS consists of the necessary edges
together with a set of redundant edges sufficient to ensure
that each necessary edge lies on some cycle in the SCSS.
We characterize the manner in which redundant edges
can lie on such cycles --- specifically, each cycle can
have at most one redundant edge and each redundant edge
lies on exactly one cycle (and thus ``provides a cycle''
for at most two necessary edges).  This allows the reduction.

A natural question is whether SCSS$_3$ is fundamentally simpler than
Bipartite Edge Cover.  In Section \ref{inv_sec} we show it is not:
\begin{theorem} \label{converse}
  Minimum Bipartite Edge Cover reduces in linear time to SCSS$_3$.
\end{theorem}

\paragraph{Comparison to undirected graphs: }
When the maximum cycle length is three,
the SCSS problem is as hard as Bipartite Matching.
When it is five, the problem is NP-hard.
When it is seventeen, the problem is MAX-SNP-hard \cite{KRY-JOURNAL}.
The latter precludes even a polynomial-time approximation scheme unless P=NP.
Thus, digraphs with bounded cycle length can have rich connectivity structure.

This highlights the fundamental difference between
connectivity in directed and undirected graphs.
The analogous problem in undirected graphs is to
find a minimum-size subset of edges preserving $2$-edge connectivity.
This problem (and many others that are NP-hard in general)
can be solved optimally in polynomial time
for graphs with bounded cycle length~\cite{ALS}.

\paragraph{Other related work: }
Moyles and Thompson \cite{MT} gave an exponential-time algorithm
for the MEG problem;
Hsu~\cite{Hsu} gave a polynomial-time algorithm for the acyclic case.

\paragraph{Contents: }
The body of the paper is organized as follows.
Section \ref{red_sec} contains the reduction of SCSS$_3$ to Edge Cover
(proving Theorem~\ref{main-result}).
Section \ref{inv_sec} notes that Edge Cover reduces in linear time to SCSS$_3$,
so that (with respect to quadratic time reductions)
the problems are equivalent.
Section \ref{appl_sec} describes the application:
the improved approximation algorithm for the general MEG problem.

\section{Reduction: SCSS$_3$ to Edge Cover} \label{red_sec}
Let $G=(V,E)$ be a strongly connected digraph with maximum cycle length $3$
or less. Assume that $G$ has at least four vertices,
none of which are cut vertices
(that is, vertices whose removal disconnects the underlying undirected graph).
This is without loss of generality,
because by standard techniques, in $O(n+m)$ time,
the cut vertices can be found
and the graph partitioned into 2-connected components.
Clearly a $c$-approximation for each component
yields a $c$-approximation for $G$.

\begin{definitions}
  An edge is {\em redundant}\/ if deleting the edge from $G$
  leaves a strongly connected graph.  Otherwise it is {\em necessary}.

  An edge $(u,v)$ is {\em unsatisfied}\/
  if there is no path from $v$ to $u$ consisting of necessary edges.

  A redundant edge $e$ {\em provides a cycle}\/ for an unsatisfied edge $(u,v)$
  if there is a path from $v$ to $u$ consisting of necessary edges and $e$.
\end{definitions}

Here is an outline of the reduction.
Since the necessary edges are in any SCSS,
the question is which redundant edges to add.
It turns out that each redundant edge lies on exactly one cycle
(Lemma \ref{property})
and thus provides a cycle for at most two unsatisfied edges.
Further, no cycle has more than one redundant edge,
so that a set of edges is an SCSS if and only if
it contains the necessary edges
and, for each unsatisfied edge $e$,
a redundant edge providing a cycle for $e$
(Lemma~\ref{properties}).

We construct an equivalent instance of Edge Cover --- an undirected graph $G'$
that has a vertex $w'$ for each unsatisfied edge $w$ in $G$
and an edge $r'$ for each redundant edge $r$ in $G$,
where $r'$ is incident to $w'$ if $r$ provides a cycle for $w$.
It turns out that the graph $G'$ is acyclic (in the undirected sense)
and thus bipartite (Lemma \ref{bipartite_lem}).

Finally, the redundant edges and the graph $G'$ can be computed
in $O(n^2)$ time (Lemmas  \ref{classifying} and \ref{building}).

\subsection{Reduction to Bipartite Edge Cover}

Here is the first essential fact:
\begin{lemma} \label{property}
  Each redundant edge lies on exactly one cycle in $G$.
\end{lemma}
\begin{proof}
  We use here the assumption that $G$ has at least four vertices,
  none of which are cut vertices.

  Since $G$ is strongly connected, each edge lies on at least one cycle.
  Suppose for contradiction that some redundant edge $(u,v)$
  lies on more than one cycle.
  There are (at least) two distinct paths from $v$ to $u$.
  At least one of the paths is of length two.
  Denote this path $(v,x,u)$.

  Since edge $(u,v)$ is redundant,
  there is a path $P_{uv}$ from $u$ to $v$ other than edge $(u,v)$.
  $P_{uv}$ must contain $x$, for otherwise $P_{uv}$ and the path $(v,x,u)$ 
  would form a cycle of more than three edges.  

  If the edge $(v,u)$ is present in $G$, then $P_{uv}$ is of length two
  (as it forms a cycle with $(v,u)$) and hence is the path $(u,x,v)$.  
  Thus, in this case,
  all six possible edges are present
  between the three vertices $u$, $x$, and $v$.
  Let $V_u$ denote the vertices reachable from $u$
  without going through $v$ or $x$.  Define $V_v$ and $V_x$ similarly.
  Using the strong connectivity of the graph and its lack of long cycles,
  one can easily show that these sets are disjoint
  and have no edges between them.
  Thus, either at least one of $u$, $x$, or $v$ is a cut vertex
  or the graph has only these three vertices.
  This contradicts our assumption about $G$.
  
  Thus, the edge $(v,u)$ is not present
  and there exists a path distinct from $(v,x,u)$ and $(v,u)$
  from $v$ to $u$.
  Denote this path, which must be of length two, by $(v,y,u)$.
  The path $P_{uv}$ must contain $y$
  for the same reason $P_{uv}$ contains $x$.  
  Thus, there is a path $Q$, without loss of generality from $y$ to $x$,
  that does not contain $u$ or $v$ (see Figure~\ref{pathBackFig}).
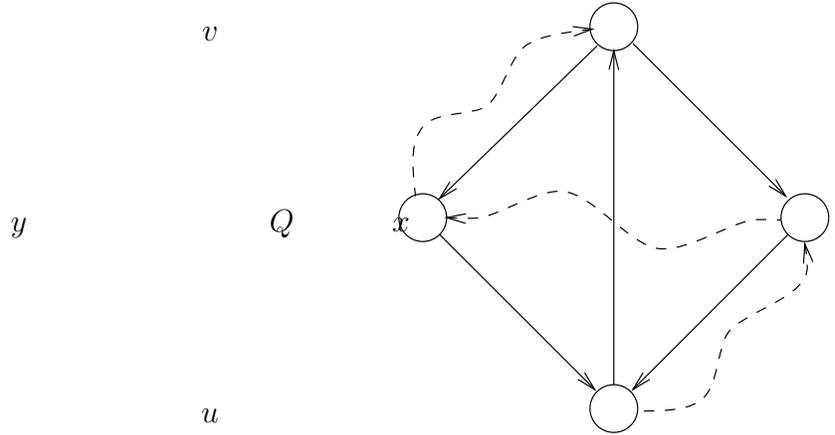
  \fig{pathBack}{pathBackFig}{No cycle has two redundant edges.}
  This is a contradiction, because the edges $(x,u)$, $(u,v)$, and $(v,y)$
  would form a cycle of length at least four with the path $Q$.
\end{proof}

\begin{lemma} \label{properties}
  A set of edges is an SCSS iff it contains the necessary edges
  and, for each unsatisfied edge $e$,
  some redundant edge providing a cycle for $e$.
\end{lemma}
\begin{proof}
  The ``if'' direction is straightforward.  To see the converse,
  first note that each cycle in $G$ contains at most one redundant edge
  (otherwise each redundant edge would lie on more than one cycle,
  violating Lemma~\ref{property}). In fact, this also implies that 
  unsatisfied edges are not redundant, otherwise we would have a cycle with
  two redundant edges.

  Since the SCSS strongly connects the graph,
  any unsatisfied edge must form a cycle with the edges in the SCSS.
  By the preceding observation, this cycle has one redundant edge.
\end{proof}

By Lemma~\ref{properties},
the problem reduces to identifying a smallest set
of redundant edges such that for each unsatisfied edge $e$ in $G$,
some redundant edge in the set provides a cycle for $e$.
By Lemma~\ref{property},
each redundant edge provides a cycle for at most two unsatisfied edges.

Build a graph $G'$ whose vertices correspond to the unsatisfied edges.
For each redundant edge $e$,
if $e$ provides a cycle for two unsatisfied edges,
add an edge between the two corresponding vertices;
if $e$ provides a cycle for one unsatisfied edge,
add a self-loop at the corresponding vertex.

By the above discussion,
a set of edges in $G'$ forms an edge cover
if and only if the corresponding set of redundant edges in the original graph,
together with the necessary edges, form an SCSS\@.

So far, we have reduced our problem to Minimum Edge Cover.
The next lemma shows that the reduction
is in fact to Minimum {\em Bipartite}\/ Edge Cover.
\begin{lemma} \label{bipartite_lem}
  $G'$ is bipartite.
\end{lemma}
\begin{proof}
  We will show that the unsatisfied edges in $G$ can be two-colored
  so that no adjacent edges have the same color.
  This gives the result as follows:  color each vertex in $G'$
  with the color of its corresponding unsatisfied edge in $G$;
  by Lemma~\ref{property},
  vertices that share an edge in $G'$ correspond to adjacent
  (and therefore differently colored) unsatisfied edges in $G$.
  Thus, no edge in $G'$ has two vertices of the same color.
  
  Assume for contradiction that the unsatisfied edges of $G$
  cannot be legally two-colored.  Then some set $C$ of the edges
  corresponds to an (odd) cycle in the underlying undirected graph.
  Let edge $(u,v)$ be one of the edges in $C$.
  We will show that there is an alternate path from $u$ to $v$,
  so that $(u,v)$ is redundant.  Since unsatisfied edges are not redundant,
  this is a contradiction.

  It suffices to show that, for each edge $(a,b)$ on $C$,
  there is a path from $b$ to $a$ that does not use $(u,v)$.
  Suppose the return path for $(a,b)$ does contain $(u,v)$.
  The return path must have length two, so either $u=b$ or $v=a$.

  We consider only the first case; the other is similar.
  Since $u=b$, this case reduces to finding a path from $u$ to $a$
  that does not use $(u,v)$, given that $(u,v,a)$ is a return path for $(a,u)$
  and that $(u,v)$ and $(a,u)$ are unsatisfied and therefore necessary.

  Suppose edge $(v,a)$ was necessary.  Then cycle $(a,u,v,a)$ would consist
  of necessary edges, so none of its edges would be unsatisfied.
  Thus, $(v,a)$ is redundant.
  Let $P_{va}$ be an alternate path from $v$ to $a$.
  $P_{va}$ must go through $u$, 
  for otherwise $P_{va}$ and the edges $(a,u)$ and $(u,v)$
  would form a cycle of length more than three.
  Thus, $P_{va}$ contains a path from $u$ to $a$ 
  that does not go through $v$.
  This portion of $P_{va}$ is the desired path.
\end{proof}

\subsection{Complexity}
To finish the proof of Theorem~\ref{main-result},
we show that the reduction can be computed in $O(n^2)$ time.

\begin{lemma} \label{classifying}
  Classifying the edges as redundant or necessary requires $O(n^2)$ time.
\end{lemma}
\begin{proof}
  Let $G$ have $n$ vertices and $m$ edges.
  Fix a root $r$ and find an incoming and an outgoing branching
  (spanning trees rooted at $r$ with all edges directed towards
  or, respectively, away from $r$).
  This can be done in $O(n+m)$ time using depth-first search.
  Let $B$ be the union of the sets of edges in the two branchings.
  There are at most $2n-2$ edges in $B$
  and the edges not in $B$ are redundant.
  This leaves $O(n)$ edges to be classified.
  Classify them using $O(n)$ time per edge as follows.
  
  Consider an edge $(u,v)$.
  Enumerate the other vertices
  to check for alternate paths from $u$ to $v$ of length two.
  If such a path exists, the edge is redundant.
  Otherwise, check for the edge $(v,u)$.
  If it exists, the edge $(u,v)$ is necessary,
  because any alternate path from $u$ to $v$ would have to have length two
  and all such paths have been checked.
  Otherwise, check all return paths of length two.
  If at least two of these paths exist,
  the edge is necessary by Lemma~\ref{property}.

  Otherwise $(u,v)$ has a unique return path $(v,w,u)$.
  If an alternate path from $u$ to $v$ exists,
  then it must use $w$ (else we get a cycle of
  length at least four).  Because of the edge $(w,u)$,
  the path from $u$ to $w$ can have length at most two.
  Similarly, the path from $w$ to $v$ can have length at most two.
  Thus, $w$ and the existence of the paths from $u$ to $w$ and $w$ to $v$
  can be determined by enumeration in $O(n)$ time.
\end{proof}

\begin{lemma} \label{building}
  Building the graph $G'$ requires $O(n^2)$ time.
\end{lemma}
\begin{proof}
  Once the edges have been classified as redundant or necessary,
  the unsatisfied edges and the return paths for the redundant edges
  can be identified in $O(n^2)$ time as follows.
  Each edge $(u,v)$ is redundant, or necessary but {\em not}\/ unsatisfied,
  if and only if it has a return path of one or two necessary edges.
  Enumerate each path of one or two necessary edges;
  let $u$ and $v$ be the first and last vertices on the path;
  if there is an edge $(v,u)$,
  then either note its return path (if it is redundant)
  or note that it is not unsatisfied (if it is necessary).
  There are $O(n^2)$ such paths, since there are $O(n)$ necessary edges.
\end{proof}

This proves Theorem~\ref{main-result}.

\section{Reduction: Edge Cover to SCSS$_3$} \label{inv_sec}
The proof of Theorem~\ref{converse} (Minimum Bipartite Edge Cover
reduces in linear time to SCSS$_3$) is somewhat simpler:

\noindent {\em Proof of Theorem~\ref{converse}. }
  Given an undirected bipartite graph,
  construct a directed graph as shown in Figure~\ref{bipartite}.
  Direct all the edges from the first part to the second part.
  Add a root vertex with edges to each vertex in the first part
  and from each vertex in the second part.
  Any edge cover in the original graph
  (together with the added edges) yields an SCSS and vice versa.
\myendproof

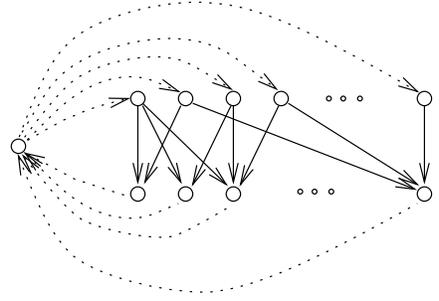
\fig{bipartite}{bipartite}{Bipartite Edge Cover reduces to SCSS$_3$.}

\section{Application to the General MEG Problem} \label{appl_sec}
Here we describe the improvement to the approximation algorithm
for the general MEG problem in \cite{KRY-JOURNAL}.
As usual, without loss of generality, assume the graph is strongly connected.
The algorithm in \cite{KRY-JOURNAL} works by repeatedly contracting cycles.
Each cycle contracted is either a longest cycle in the current graph,
or has length at least some constant $k$.
The set of contracted edges yields the set $S$.
As $k$ grows, the performance guarantee of the algorithm
tends rapidly to $\pi^2/6\approx 1.64$.

Assume $k \ge 4$.
Modify the algorithm so that as soon as the current graph
has maximum cycle size three or less,
it solves the problem optimally (using Theorem~\ref{main-result})
and returns the edges in the solution for the current graph
together with the edges on previously contracted cycles.

To {\em contract}\/ an edge is to identify its endpoints in the graph
as a single vertex;
to contract a cycle is to identify all vertices on the cycle.
We use the following result from \cite{KRY-JOURNAL}:
\begin{proposition}[\cite{KRY-JOURNAL}] \label{lower-bound}
  If the maximum cycle length in an $n$-vertex graph is $\ell$,
  then any SCSS has at least $(n-1)\ell/(\ell-1)$ edges.
\end{proposition}
The proof is that any strongly connected graph can be contracted
to a single vertex by repeatedly contracting cycles whose edges
are in the SCSS; the ratio of edges contracted to vertices lost
when one of these cycles is contracted is at least $\ell/(\ell-1)$.

{\em Proof of Corollary~\ref{corollary}.}\/
{%
  Initially, let the graph have $n$ vertices.
  Assume $n_i$ vertices remain in the contracted graph
  after contracting cycles with $i$ or more edges ($i=k,k-1,\ldots,4$).  
  Finally, we get a graph $H$ (with $n_4$ vertices) that has no cycles of
  length four or more;
  the algorithm solves the SCSS problem for $H$ optimally.

  How many edges are returned?
  Let $\Opt{G}$ denote the minimum size of an SCSS of $G$.
  In contracting cycles with at least $k$ edges,
  at most $\frac{k}{k-1}(n-n_k)$ edges are contributed to the solution.
  For $4 \le i<k$, in contracting cycles with $i$ edges,
  $\frac{i}{i-1}(n_{i+1}-n_i)$ edges are contributed.
  The number of edges returned is thus at most
  \begin{displaymath}
    \frac{k}{k-1}(n-n_k)
    + \sum_{i=4}^{k-1}\frac{i}{i-1}(n_{i+1}-n_i)
    + \Opt{H}.
  \end{displaymath}
  A little work shows this is equal to
  \begin{displaymath}
    \frac{k}{k-1}(n-1)
    + \sum_{i=5}^k \frac{n_i-1}{(i-1)(i-2)}
    - \frac{4}{3} (n_4-1)
    + \Opt{H}.
  \end{displaymath}
  Since $\Opt{H} \le 2(n_4-1)$, substituting for $n_4$ gives the upper bound
  \begin{displaymath}
    \frac{k}{k-1}(n-1)
    + \sum_{i=5}^k \frac{n_i-1}{(i-1)(i-2)}
    + \frac{1}{3} \Opt{H}.
  \end{displaymath}
  Clearly $\Opt{G} > n-1$.  For $4\le i \le k$,
  when $n_i$ vertices remain, no cycle has more than $i-1$ edges.
  By Proposition~\ref{lower-bound},
  any SCSS of the current graph (and therefore any SCSS of $G$)
  has at least $(n_i-1)(i-1)/(i-2)$ edges.
  Also $\Opt{G}\ge\Opt{H}$.
  Using these three facts, the above quantity, divided by \Opt{G}, is less than
  \begin{eqnarray*}
    \lefteqn{\frac{k}{k-1}
      + \sum_{i=5}^k \frac{1}{(i-1)(i-1)}
      + \frac{1}{3}}
    \\ &=&
        \frac{1}{k-1} + \sum_{i=1}^{k-1} \frac{1}{i^2} -\frac{1}{36}.
  \end{eqnarray*}
  Using the identity (from \cite[p.75]{Knuth1}) 
  $\sum_{i=1}^\infty \frac{1}{i^2} = \frac{\pi^2}{6}$,
  this is equal to
  \begin{eqnarray*}
    \lefteqn{
      \frac{\pi^2}{6} - \frac{1}{36} +\frac{1}{k-1}
      - \sum_{i=k}^\infty \frac{1}{i^2}}
    \\&\le& \frac{\pi^2}{6} -\frac{1}{36}+ \frac{1}{k-1} 
    - \sum_{i=k}^\infty \frac{1}{i\,(i+1)}
    \\&=& \frac{\pi^2}{6}-\frac{1}{36} + \frac{1}{k-1} - \frac{1}{k}
    \\&=& \frac{\pi^2}{6}-\frac{1}{36} + \frac{1}{k(k-1)}.
  \end{eqnarray*}
  }
\myendproof

Similarly to \cite{KRY-JOURNAL},
standard techniques can yield more accurate estimates,
e.g.,
\(\frac{\pi^2}{6} -\frac{1}{36}
+ \frac{1}{2k^2} + O\left(\frac{1}{k^3}\right).\)
Also following \cite{KRY-JOURNAL},
if the graph initially has no cycle longer than $\ell$ ($\ell \ge k$),
then the analysis can be generalized to show a performance guarantee
of $\frac{k^{-1}-\ell^{-1}}{1-k^{-1}}
+\sum_{i=1}^{k-1}\frac{1}{i^2} - \frac{1}{36}$.
For instance, in a graph with no cycle longer than $5$,
the analysis bounds the performance guarantee (when $k=5$) by $1.396$.

\paragraph{Acknowledgments: }
We thank R.~Ravi and Klaus Truemper for helpful discussions.
We also thank the referees for useful comments on a preliminary draft
of this paper.

\end{document}